\documentclass{article}
\pdfoutput=1
\usepackage{spadre2008}
\usepackage{graphicx}
\frompage{000} \topage{000}                                              
\def\rightmark{Conical Di-jet Correlations in AdS/CFT}
\def\leftmark{M. Gyulassy et.al.}

\title{Conical Di-jet Correlations from a Chromo-Viscous Neck in AdS/CFT}
\authors{
{M.\ Gyulassy$^1$, J.\ Noronha$^1$, and G.\ Torrieri$^2$%
}\\[2.812mm]
{\normalsize
\hspace*{-8pt}$^1$Department of Physics, Columbia University, 538 West 120$^{th}$ St., New York, NY 10027, USA\\
\hspace*{-8pt}$^2$Institut f\"ur Theoretische Physik and FIAS, J.W. Goethe Universit\"at, Frankfurt, Germany
}}

\abstract{
We show that Mach-like
correlations from heavy quark jets
in the Gubser et al AdS/CFT string drag model
arise not from the expected (weak) hydrodynamical Mach sound wakes zone
but from a near ``Neck'' zone where strong
chromo-viscous field effects
 dominated and lead to a distinctive stress in
the plasma of the form
$T^{\mu\nu}_{Neck} \propto \surd \lambda T_0^2 \,Y^{\mu\nu}(x)/x^2$.
We propose that measurement of the jet velocity dependence of moderate $p_T$
hadrons associated with future
{\em identified heavy quark jets} at RHIC and LHC
could be used to look for this novel source of conical correlations
which are unrelated to Mach's law.}


\begin{document}

\maketitle
\setcounter{page}{1}

\section{Introduction}\label{intro}
The observation of Mach-like conical correlations reported from the Relativistic Heavy Ion Collider (RHIC) between
tagged jets and associated moderate $p_\perp$ hadrons \cite{Adler:2005ee} has been interpreted \cite{Stoecker:2004qu,shuryakcone}
as providing additional evidence for fast relaxation time and the near perfect fluid property of strongly coupled quark gluon plasmas (sQGP) \cite{Gyulassy:2004zy}. This has generated wide interest because if  Mach's law were valid, $\cos\theta_M= c_s/v$, the correlation pattern could provide a direct measurement of of the time averaged speed of sound in this new state of matter.

The question addressed in this talk is whether the perfect fluid hydrodynamic
interpretation via Mach's law is unique
or could there be other novel nonequilibrium dynamical
mechanisms involved.  We do not consider here the possibility that the
experimental ZYAM background subtraction method could be partly responsible
for the observed away side dip  (see \cite{yZYAMunhappy}).
Nor do we consider more conventional
sources correlations such as those due to jets deflected
by the strong radial flow in A+A \cite{Pruneau:2007ua}.  Rather, we consider
the possibility \cite{gubsermach,Herzog:2006gh,NGTnonmach} that di-jet correlations
are connected to Anti-de Sitter/Conformal Field Theory (AdS/CFT) \cite{maldacena} models that have been proposed to explain some of the surprising sQGP
properties and dynamics. Supergravity models provided a new way of understanding the unexpected ``$\frac{3}{4}\times$SB'' entropy law \cite{Gubser:1996de} that appears to explain numerical lattice QCD data above $T_c$ and the unexpected low viscosity of SQGP \cite{Son:2007vk} determined from collective elliptic flow data. In addition, these models \cite{gubsermach,Herzog:2006gh} offer an alternate to the pQCD's way \cite{Djordjevic:2005db} of explaining the high opacity of sQGP inferred from heavy quark jet quenching. Therefore, it is of interest to look for further observables to test the developing AdS/CFT phenomenology via high energy nuclear collisions. The recent di-jet correlation data from RHIC and the future data with identified heavy quark jets at RHIC and LHC offer an attractive possibility.

We discuss here results, first presented at WWND08 and reported in \cite{NGTnonmach}, based on a  Cooper-Frye freeze-out analysis of the AdS string drag model stress solutions obtained by Gubser, Pufu and Yarom \cite{gubsermach}. We show that away trigger side double-shoulder (Mach-like conical)  azimuthal correlation arise in this AdS string drag model not from the Mach wake zone but from a novel chromo-viscous ``Neck'' region close to the quark. We prove that in the supergravity limit, only the singular Neck zone where strong external fields seem to drive an explosive transverse collective flow that can give rise to conical correlations. We propose that this novel mechanism can be tested experimentally by observing a nearly heavy quark jet velocity independence of the conical correlations in contrast to Mach's law \cite{NGTnonmach}.

The supergravity string drag model \cite{Herzog:2006gh} based on the AdS/CFT correspondence reveal (as shown in Fig 1) a rather complex pattern of energy-momentum stress, $T^{\mu\nu}(x)$, perturbations in the wake of a supersonic string moving above a blackbrane in an AdS$_5\times$S$_5$
curved background. The idea is that these solutions provide a holographic image of the stress induced by a heavy quark jet in a very strongly coupled conformal supersymmetric  Yang-Mills plasma (SYM) cousin of sQGP/QCD. While strings moving in a 10D gedanken curved bulk seem far removed from the real world, experimentalist at RHIC \cite{Xu:2006dx} and LHC \cite{Carminati:2004fp} can soon test or falsify these ideas by measuring  identified heavy quark b/c jet quenching \cite{Horowitz:2007su} and di-jet correlations \cite{NGTnonmach}.

\section{String Drag Stress Zones}
The energy-momentum stress induced by a heavy quark moving
at velocity $v$ through a supersymmetric Yang-Mills plasma (SYM)
can be decomposed into terms that dominate in different spatial
regions or zones. In  the supergravity
$N_c\gg 1,\, g_{YM}^2\ll 1,\, \lambda=g_{YM}^2 N_c\gg 1$ limit the AdS Minkowski boundary stress can be written as
\begin{equation}
T^{\mu\nu}(x)=T^{\mu\nu}_0+\delta T^{\mu\nu}_{Mach}
+ \delta T^{\mu\nu}_{Neck}+ \delta T^{\mu\nu}_{Coul}.
\end{equation}
We choose coordinates $x_1=z-vt$ along the direction,
$n^\mu=(0,-1,0,0)$, of the away side jet,
and $x_\perp$ as the cylindrical transverse cylindrical
radial coordinate perpendicular to the jet axis.
In this system the beam direction is $(0,0,0,1)$ and the trigger jet
is in direction $-n^\mu$.

The far zone ``Mach'' part
of the stress  can be expressed in terms the local
temperature and fluid (Landau) flow velocity
fields $(T(x_1,x_\perp), U^\alpha(x_1,x_\perp))$
through the Navier-Stokes stress form
\begin{equation}
\delta T_{Mach}(x_1,x_\perp)= \frac{3}{4}
K\left\{T^4\left(\frac{4}{3}U^\mu U^\nu-\frac{1}{3}g^{\mu\nu}
+ \frac{\eta}{sT} \partial^{\{ \mu}U^{\nu\}} \right)- T^{\mu\nu}_0\right\}
\theta(1-3Kn)
\end{equation}
where in the supergravity limit $\eta/s=1/4\pi$ \cite{Son:2007vk}, $ \partial^{\{ \mu}U^{\nu\}} $ is the symmetrized traceless
shear flow velocity gradient, and $Kn$ is the local Knudsen number defined below.

Currently only drag solutions in a static, uniform
$T=T_0$ SYM have been tabulated \cite{gubsermach}.
The background ${\cal N}=4$ infinitely coupled SYM plasma
(in the strict $\lambda=\infty$ limit)
is characterized by the remarkably simple equation of state
 ``3/4'' of the ideal conformal Stefan-Boltzmann form:
$T^{\mu\nu}_{0}=P_0\, diag(3,1,1,1)$, with $P_0=\epsilon_{SYM}/{3}$ and
$\epsilon_{SYM}=\frac{3}{4} KT_0^4$.
The Stefan Boltzmann constant is $K=(N_c^2-1)\pi^2/2$ for this imaginary
world of $N_c^2\rightarrow\infty$ massless adjoint color
vector fields, fermions, and bosonic degrees of freedom. This make believe gedanken
world offers great theoretical calculational
advantages due to its very high SO(4,2) isometry constraints
corresponding to conformal as well as Poincare invariance on the gauge theory
side.

No one should be squirmish about jumping into the imaginary AdS$_5$ black hole -
after all, quarks and gluons, the sQGP, imaginary time partition functions, lattice QCD,
pomerons cuts in the complex angular momentum plane, $x=0$ CGC
and all other ``standard model'' concepts
 are simply symbols for specific calculational algorithms invented to
reduce data to as few parameters as possible
and to predict new falsifiable phenomena.
AdS/CFT thus  viewed is just a novel set of algorithmic tools
with new calculus tricks that have been mostly developed thus far to try
to predict strongly coupled non-Abelian gauge dynamics phenomena.
Identified heavy quark dijet tomography, among other observables
such as direct photons, leptons and hyperons produced in A+A
at RHIC and LHC, will test which algorithm most efficiently
captures the  experimental constraints:
either the downward extrapolation $3 \leftarrow N_c \ll\infty,
\alpha_c \leftarrow \lambda/12\pi \ll \infty$
of the AdS supergravity calculus
or the upward extrapolation $0\ll \alpha_{YM}=\lambda/12\pi \rightarrow
\alpha_c$ of the pQCD calculus.  Here,
$\alpha_c\approx 0.5$ is Gribov's critical strong QCD
coupling \cite{Dokshitzer:2004ie} proposed to explain
the mystery of confinement.
Recent progress to solve numerically
radiative pQCD transport theory \cite{Xu:2007ns}
showed that $\eta/s$ apparently saturates
the AdS/CFT KSS bound \cite{Son:2007vk} when $\alpha_s$ is extrapolated
near $\alpha_c$.

The apparent complementarity of both approaches when extrapolated near to the critical coupling scale suggests that
\begin{equation}
\lim_{\alpha_s\rightarrow \alpha_c^-} \; pQCD \approx sQGP \approx
\lim_{\alpha_c^+\leftarrow \lambda/12\pi} \; \lim_{3^+\leftarrow N_c} \; AdS/CFT .
\label{mysqgpeq}
\end{equation}
 It is generally
 believed that static
equilibrium properties of sQGP are reliably predicted by lattice QCD
and at best AdS/CFT can only provide analytic insight into what went
on during $10^{18}$ flops in the computer.
Eq.(\ref{mysqgpeq}) is however proposed to be
relevant to nonequilibium
real time dynamical processes as well for which LQCD
provides no information (see, however, the recent results \cite{Meyer:2007ic}).
Future experiments at RHIC and LHC are needed to answer
how large a domain of observables and in which corners of phase
space Eq.\ (\ref{mysqgpeq}) may hold. On the theory side, further advances in
both pQCD radiative transport theory and in the AdS phenomenology
will also be essential to improve and quantify the level of adequacy
denoted by $\approx$ symbol.

The theta function, $\theta(1-3 Kn)$ defines
a  far ``Mach'' zone where
the equilibration rate
is large  enough compared to the stress gradient scales
that  Navier-Stokes dissipative hydrodynamics
provides an adequate description
of the evolution. Numerical  transport theory results indicate
that local equilibrium is achieved when the effective number of collisions $1/Kn$ exceeds about $3$.
Operationally, we define \cite{Noronha:2007xe}  the Knudsen number field
$Kn(x) = \Gamma_s |\nabla\cdot\vec{S}|/|\vec{S}|$
in terms of the
$S^i=T^{0i}$ Poynting vector field, and the sound attenuation length
$\Gamma_s\equiv 4\eta/3Ts \ge
1/3\pi T_0$ that
is  bounded from below in ultra-relativistic systems by
the uncertainty principle \cite{Danielewicz:1984ww}. In the conformal
supergravity limit $\Gamma_s$ saturates the KSS bound \cite{Son:2007vk}
$1/3\pi T_0$. We note that the ``Mach'' zone of of the AdS solution
contains  not only the minimally diffused  conical ``Mach'' sound wake,
but also a near axis trailing diffusion plume wave \cite{shuryakcone,NGTnonmach} seen in Fig.1.

The far zone excludes a compact
Neck zone (see Fig.1 insert) close to the heavy quark jet
where  the local Knudsen number is large  and the Navier-Stokes
form of the stress is not applicable.
In the Neck the effective number of collisions over the
scale of stress variation, $1/Kn(x)<3$,
and even uncertainty bounded equilibration rates are too
small to maintain local equilibrium. As shown in
\cite{Noronha:2007xe,Yarom:2007ni} the non-equilibrium zone is characterized
by a stress of the form
\begin{equation}
\delta T_{Neck}(x_1,x_\perp)\approx \theta(3Kn(x)-1)
\frac{\surd \lambda T_0^2}{x_\perp^2 +\gamma^2 x_1^2}Y^{\mu\nu}(x_1,x_\perp)
\end{equation}
where $Y^{\mu\nu}$ is a dimensionless ``angular'' tensor
field. At very small distances from the jet,
$Y^{\mu\nu}$ reduces to the analytic
analytic Yarom stress tensor,
$(x_\perp^2 +\gamma^2 x_1^2)T^{\mu\nu}_{Y}(x_1,x_\perp)$ \cite{Yarom:2007ni,Gubser:2007nd}.  Fig.2 shows $T^{00}_Y/\epsilon_{SYM}$
and its Knudsen field. Note that
the $Kn(x)=1/3$ and the $T^{00}_Y/\epsilon_{SYM}=1/3$
contours are similar for this velocity.
They are also similar in the range $c_s< v< 0.9$,
where we expect the Mach angle to vary the most. For
$v\rightarrow 1$, where the Mach angle saturates at
$\theta_M=acos(1/\surd 3)=0.955$ rad,
the $Kn$ distribution acquires an additional distinct
Lorentz contracted pancake component between the two lobes
shown for $v=0.9$ in Fig.2.

Within the Neck zone, there is also an inner ``Head'' region
where the stress becomes dominated by the contracted
Coulomb self field stress of the quark shown in Fig.3.
The Head zone can formally  be defined as in
Ref. \cite{Dominguez:2008vd} by equating the analytic
Coulomb energy density \cite{gubsermach,Friess:2006fk}, $\varepsilon_C(x_{1},x_{\perp})$,
to the analytic near zone Yarom energy density
\cite{Yarom:2007ni}, $\varepsilon_Y(x_{1},x_{\perp})$.
The Head zone is  a Lorentz contracted
pancake with longitudinal thickness $\Delta x_{1,C} \,\pi\, T_0 \sim 1/\gamma^{3/2}$
and an effective transverse radius
 $ \Delta x_{\perp,C}\,\pi\, T_0 \sim 1/\gamma^{1/2}$, which is in agreement with the general considerations in
Ref. \cite{Dominguez:2008vd}.  However,
the numerical results reported  in \cite{Noronha:2007xe}  show
that for $v=0.99$ jets, the Knudsen Neck has a two lobe structure
also seen in Fig 2 for $v=0.9$. The lobes are  nearly independent of $v$ and
the lobe region thickness is
$\Delta x_{1,Kn} \sim 1/\pi T_0\gg \Delta x_{1,C}$.
The second thin pancake component of the $Kn$ that develops for large $\gamma$
(not shown) is similar to the shape of the Head zone.
The relative independence of the two lobe component
of the Red Neck zone on $v$ is
in agreement with the parametric dependence $\Delta x_{1,N}\propto 2/\pi T_0 \sim 6\Gamma_s$ expected from relativistic
uncertainty principle bounded dissipation rates
\cite{Son:2007vk,Danielewicz:1984ww}

The rich and non-intuitive structure of the near zone
$O(\surd\lambda T^2/x^2)$ stress is presumably associated
with the non-trivial way in which the Coulomb self field of the jet
couples to the viscous but also conductive SYM fluid. Whether the AdS large chromo-viscous dynamics
in the Neck/Head region is a nonperturbative generalization
of QCD chromo-viscous-hydrodynamics \cite{Heinz:1985qe,Selikhov:1993ns} is an interesting open question.

\section{Cooper-Fryed AdS Holograms}
We now turn to the observable consequences
of the AdS string drag stress model
by assuming a Cooper-Frye (CF) hadronization scheme \cite{Cooper:1974mv}
as in \cite{shuryakcone,heinzcone,Betz:2008js}.
This is an strong model assumption
on top of the AdS calculus and will need much closer scrutiny
in the future. We present it as a first attempt to try to
map the AdS boundary stress (the hologram) into hadronic observables.

The axial symmetry with respect to the trigger jet axis
allows us to write
 $U^{\mu}(x_1,x_\perp)= ( U^0, U_1, U_{\perp}\cos\varphi, U_\perp\sin\varphi)$. The CF associated away side azimuthal distribution of massless ($\sim$ pions)
at midrapidity,
$f(\phi)=dN/(2\pi p_{T}^2)dp_{T}dyd\phi |_{y=0}$, with respect to the nuclear beam axis is given by \cite{NGTnonmach}
\begin{equation}
f(\phi;V)=
\int_{V} dx_1 dx_\perp x_\perp 
 \left(
e^{-\frac{p_T}{T}\left[U_0-U_1\cos(\pi-\phi)\right]} I_0(a_\perp)-e^{-p_T/T_0} \right)
\end{equation}
where $V$ is a particular (Mach, Neck, or Head) zone volume of interest
and $a_\perp=p_\perp U_\perp\sin(\pi-\phi)/T$ and $I_0$ is the modified Bessel function.

In the supergravity approximation $a_{\perp} \sim \mathcal{O}\left(\frac{\sqrt\lambda}{N_{c}^2}\right) \ll 1$ and, thus, we can expand the  Bessel function
to get the approximate equation for the distribution
\begin{equation}
f(\phi) \simeq e^{-p_{T}/T_0}\frac{p_{T}}{T_0}
\left[\frac{\langle \Delta T\rangle}{T_0} + \langle U_{1}\rangle \cos
(\pi-\phi) 
\right]\label{expandedCFfinal}
\end{equation}
where
 deviations from isotropy are then controlled by the following global moments $\langle \Delta T\rangle= \int_{V} dx_1 dx_\perp x_\perp \,\Delta T$ and $\langle U_1\rangle= \int_{V} dx_1 dx_\perp x_\perp \,U_1$.
Therefore, in the strict supergravity
 $N_c\rightarrow \infty $ limit,  $\Delta T/ T_0\ll 1$ and $\vec{p}\cdot\vec{U}/T\ll 1$ with tiny corrections $\mathcal{O} (\lambda/ N_{c}^4)$
and formally for any fixed $p_T/T_0$ associated hadron
the away side distribution is a trivial broad peak about $\phi=\pi$
regardless of the nice Mach wake evident in Fig. 1.
This simple theorem holds independent
of the strength of the diffusion wake \cite{shuryakcone,gubsermach,Gubser:2008vz} formed behind the heavy quark. It holds for any $\Delta T(x)$ and $U(x)$
fields as long as they are small.

The only way that a nontrivial angular correlation can arise in the
AdS/CFT string drag model
 is if we relax the formal $N_c,\,\lambda\to\infty$ but $\sqrt{\lambda}/N_c^2\to 0$
condition used to derive the stress and boldly extrapolate, in the
sense of Eq.(3),  toward  more ``physical''
parameters to  make contact with our QCD world.

\section{Numerical Results}
We computed $f(\phi;V)$ with
$N_c=3, \lambda=5.5$
\cite{gubserlambda}  for $v=0.58,0.75,0.90$,
in a static uniform background
from the tables of $T^{00}$ and $T^{0i}$ provided by Gubser et al \cite{gubsermach}. Our total CF volume is defined by $-14<X_1\,(\pi T_0)<1$, $0<X_{\perp}\,(\pi T_0)<14$, and $\varphi\in [0,2\pi]$. Here we define the head of the jet as the volume where $\xi>0.3$, which roughly corresponds to the region between $-1<X_1\,(\pi T_0)<1$ and $0<X_{\perp}\,(\pi T_0)<2$. We show results for the
azimuthal angular correlations in Fig.4. The blue curves exclude the chromo-viscous Neck zone from the CF volume. On the other hand, the red Neck
curves only include the Neck zone approximated here by $\delta T^{00}(x) > 0.3\,\epsilon_{SYM}$. Only the red Neck curves display the double-peak structure,
while the ``Mach'' zone is too weak even in the $N_c=3$ extrapolation of AdS
to produce a dip at $\phi=\pi$.. For $v=0.9$ the two peaks from the Neck zone appear at angles accidentally similar to
the putative Mach cone angle.
However, the surprising result shown for $v=0.58$ and $v=0.75$
indicates that unlike hydrodynamic Mach cones, the double peaks are
relatively independent
of $v$, in violation Mach's law indicated by the small arrows.
We propose that looking for deviations from Mach's law
for supersonic but not ultrarelativistic {\em identified heavy quark}
 jets could test this novel prediction of the AdS/CFT drag model.

\underline{Acknowledgments:}

We thank S. Gubser, S. Pufu, and A. Yarom for providing
their numerical stress tables and B. Betz and H. St\"ocker
 for useful discussions. J.N. and M.G. acknowledge
support from DOE under Grant No. DE-FG02-93ER40764. M.G. thanks the support from DFG, ITP, and FIAS at J.W. Goethe University. G.T. thanks the Alexander Von Humboldt foundation and J.W. Goethe University for support.


\begin{figure}[bht]
\centerline
%
\centering
{\includegraphics[width=4.in]{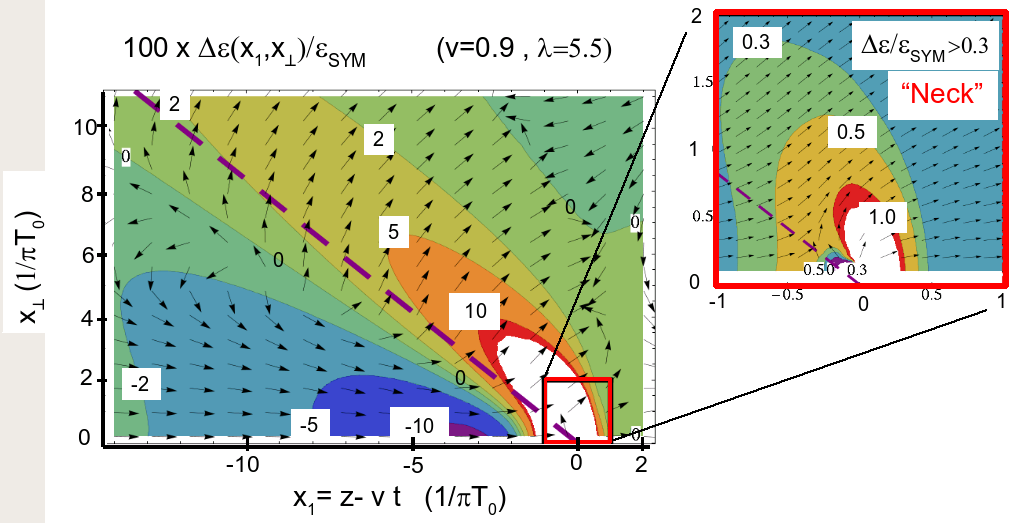}}
\hfill
{\includegraphics[width=4.in]{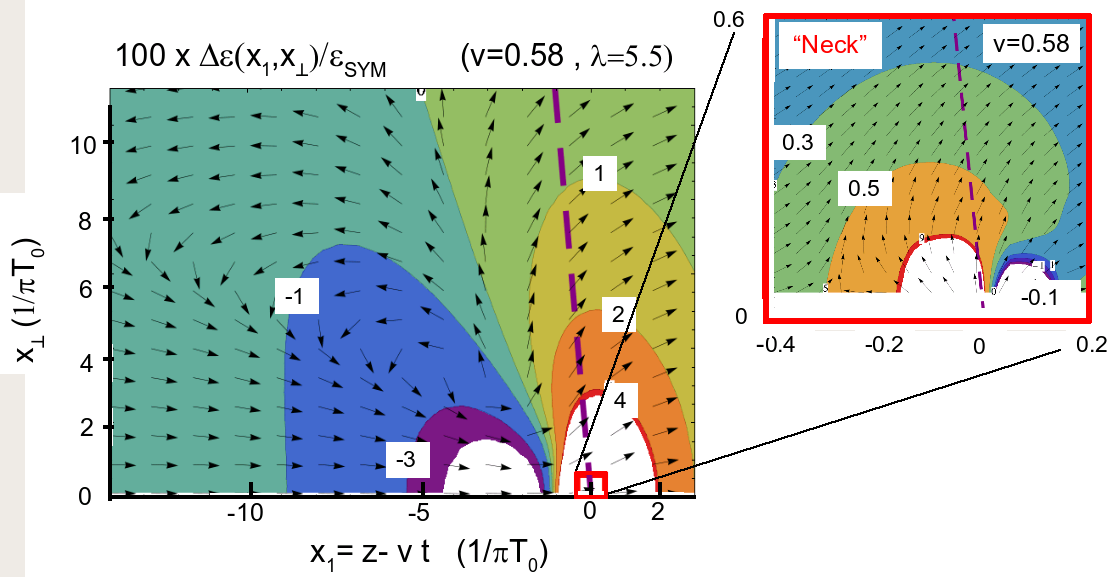}}

%
\caption{\label{Stress}
The relative energy density perturbation $\Delta\epsilon(z-vt,r)/\epsilon_{SYM}$ due to a heavy supersonic quark jet with $v/c=0.9$ top and $v=0.58>1/\surd 3$.
 The numerical results are from Gubser et al
\ {\protect\cite{gubsermach}} for using the AdS/CFT string drag calculus
for a $\mathcal{N}=4$ SYM plasma for $N_c=3,\;\lambda=g^2_{YM},\,N_c=5.5$. Left panels show the far zone (amplified by a factor of 100). The Mach wake zone is above the dashed Mach cone line, $\cos\phi_M=1/(\surd 3 v)$, while the Diffusion plume zone lies below that line. The normalized momentum flux flow directions are indicated by arrows. The insert shows the nonequilibrium AdS Neck zone (with Coulomb stress subtracted for clarity). The Neck here is
 defined approximately by $\Delta\epsilon/\epsilon_{SYM}>0.3$. Strong transverse energy flow is generated near the jet.}
\end{figure}
\end{center}

\begin{center}
\begin{figure}[th]
\centerline
%
\centering
{\includegraphics[width=4.in]{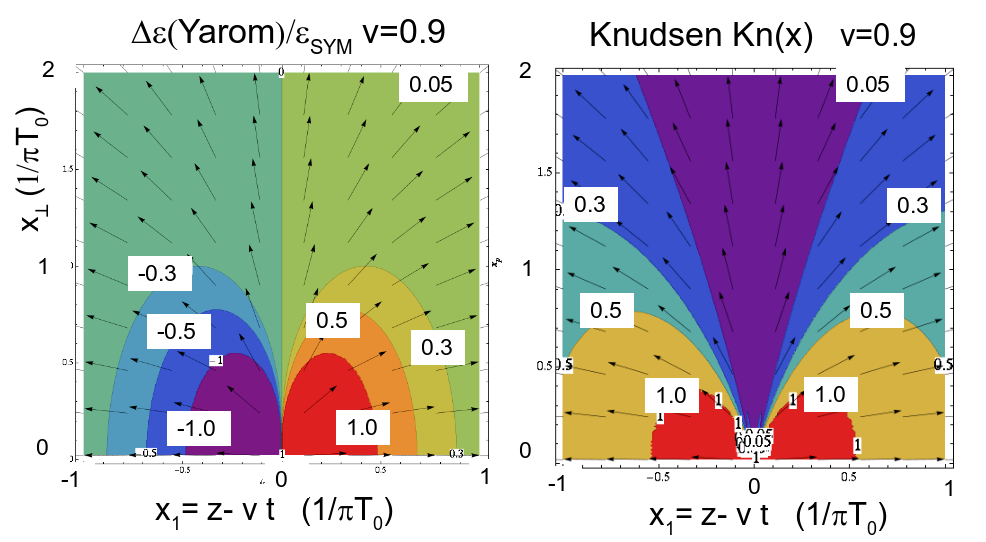}}

%
\caption{\label{knplot}
The Neck zone fractional energy density perturbation
$\Delta\epsilon(Yarom)/\epsilon_{SYM}$ for $v/c=0.9$
using the analytic near field stress from Ref.{\protect\cite{Yarom:2007ni}}.
The higher temperature (red) $z>vt$ Neck zone
with positive $\Delta\epsilon>0$
contributes more to the correlation signature
than the cooler
(blue) $z<vt$  Neck zone.
The local Knudsen number field for the
near zone Yarom stress is shown on right panel.
Note that the $3Kn>1$ region defining the  Knudsen Neck Zone
 is similar to the region defined by large
energy density perturbation $\Delta \epsilon/\epsilon_{SYM}>0.3$.
}
\end{figure}
\end{center}

\begin{center}
\begin{figure}[th]
\centerline
%
\centering
{\includegraphics[width=4.in]{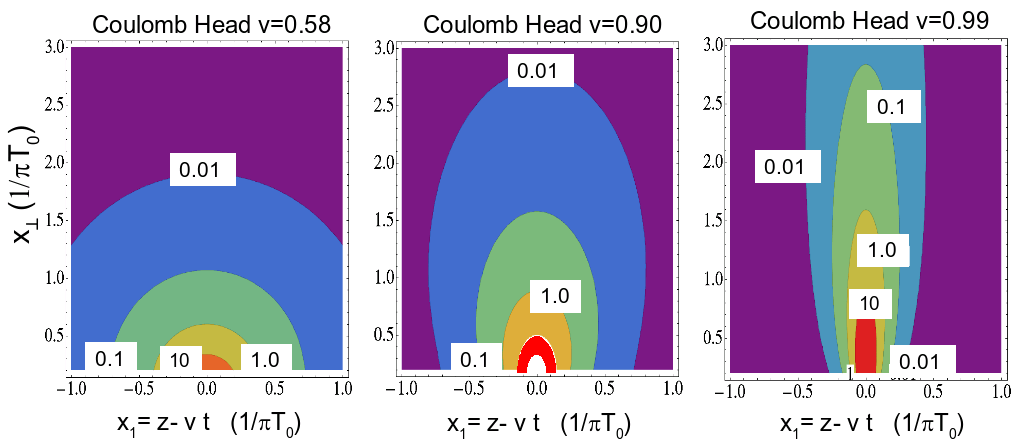}}

%
\caption{\label{Coulplt}
The logarithmic contours of the
fractional energy density $\Delta \epsilon_{Coul}/\epsilon_{SYM}$
of  the Coulomb Head  are compared for $v=0.58, 0.90,0.99$ cases.
Ohm's law generates a chromo current through the plasma mostly transverse
to the jet as $v\rightarrow 1$ and leads via Joule heating
to high temperatures/pressures that produce plasma
flow transverse to the jet. This observed conical correlation
is thus unrelated to Mach's law in the AdS model.}
\end{figure}
\end{center}
\begin{center}
\begin{figure}[ht]
\centerline
%
{\includegraphics[width=4.6in]{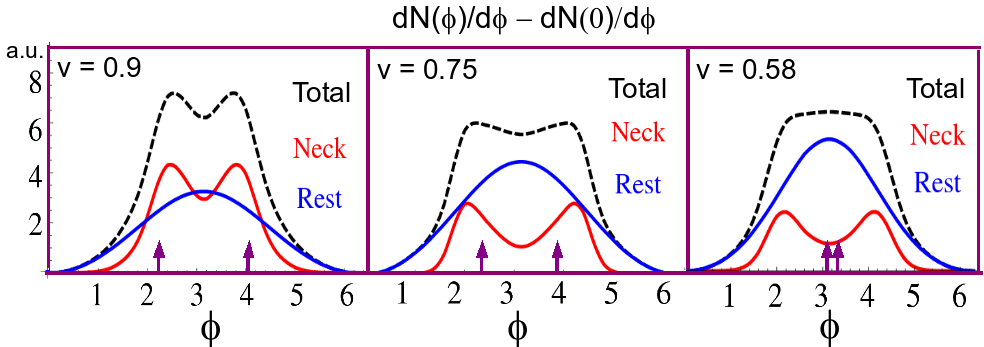}}
%
\caption{\label{angulardistplot}
Midrapidity azimuthal away side associated angular
distribution (scaled in arbitrary units, a.u.)
from the Cooper-Frye freezeout of the AdS/CFT string drag model $(T(x),\vec{U}(x))$ fields from \cite{gubsermach}.
 Three cases for various heavy quark jet velocity
and associated hadron transverse momentum ranges,
$1: (v/c=0.9, p_T/\pi T_0=4$-$5$, $2: (v/c=0.75, p_T/\pi T_0=5$-$6$), and $3: (v/c=0.58, p_T/\pi T_0=6$-$7$), are compared. The short arrows show the
expected Mach angles. The red curves showing the double shoulder away side dip (conical) correlations are from the Neck region defined here
as where $\Delta\epsilon/\epsilon_{SYM}>0.3$ (see fig 1).
The blue curves result from integrating only in the far Mach zone outside
the Neck region and show no sign of the weak Mach wake seen in Fig.1 because
to the NO-GO freezeout theorem Eq.(6) remains in force even for our
$N_c=3,\lambda=5.5$ downward extrapolation from the supergravity limit.
The sum total correlation exhibits a double shoulder correlation for $v<0.9$
arising from the chromo-viscous near zone that is however
unrelated to the Mach wakes seen in Fig.1. }
\end{figure}
\end{center}
\vspace{-0.35in}

\end{document}